\documentclass{article}

\usepackage{graphicx}
\usepackage{cite}

\def\a{\alpha}
\def\b{\beta}
\def\ga{\gamma}
\def\de{\delta}   %% NON ridefinire come \d !!!!
\def\eps{\varepsilon}
\def\vphi{\varphi}
\def\la{\lambda}

\def\s{\sigma}

\def\vphi{\varphi}

\def\P{\mathcal{P}}

\def\pa{\partial}

\def\o+{\oplus}

\def\<{\langle}
\def\>{\rangle}

\def\({\left(}
\def\){\right)}
\def\=#1{\bar #1}
\def\~#1{\widetilde #1}
\def\wt#1{\widetilde #1}
\def\.#1{\dot #1}
\def\^#1{\widehat #1}

\def\"#1{\ddot #1}

\def\eeq{\end{equation}}
\def\beq{\begin{equation}}

\def\beql#1{\begin{equation} \label{#1}}

\def\eqref#1{(\ref{#1})}

\def\symmref{AVL,CGbook,KrV,Olver1,Olver2,Stephani}
\def\sderef{Arnold,Evans,Fre,Ikeda,Kampen,Oksendal,Stroock}
\def\stochsymmref{GRQ1,GRQ2,Unal,SMS,Koz1,Koz2,Koz3,GS17,GGPR,GL1,GL2,Koz18a,Koz18b,KozB,GLS,GW18}
\def\appref{Kamenev,Ai,Band,Fia,Barb,Cogn,LW1,LW2,Mao,Pro,Spa,SFV}
\def\ranref{JSL,Nas,Nas2,NFQ,Ova,Pas}

\begin{document}

\title{Integration of the stochastic logistic equation via symmetry analysis}

\author{Giuseppe Gaeta \\ {\it Dipartimento di Matematica, Universit\`a degli Studi di Milano} \\ 
{\it via Saldini 50, 20133 Milano (Italy)} \\ {\tt and} \\ {\it SMRI, 00058 Santa Marinella (Italy)} \\
{\tt giuseppe.gaeta@unimi.it}}

%\begin{document}

\maketitle

\begin{abstract}

\noindent We apply the recently developed theory of symmetry of
stochastic differential equations to a stochastic version of the
logistic equation, obtaining an explicit integration, i.e. an
explicit formula for the process in terms of any single
realization of the driving Wiener process.

\end{abstract}

\section{Introduction}

The logistic equation \beql{eq:Logistic} \frac{dx}{dt} \ = \ \a \,
x \ - \ \b \, x^2 \eeq (with $\a , \b$ positive real constants) is
ubiquitous whenever we have a saturated growth and is thus a
fundamental equation in Mathematical Biology and in other
contexts, e.g. Chemical Physics (law of mass action). By a
standard linear change of variables this can always be taken to
the standard form \beql{eq:LogSta} dx/dt \ = \ A \ x \ ( 1 \, - \,
x ) \ ; \eeq we will however keep to the general form
\eqref{eq:Logistic} so that $x$ retains its original meaning. The
logistic equation is of course readily integrated by separation of
variables.

In applications, $x(t)$ often represents some population and $x_*
= \a/\b$ a limit level for such a population, e.g. a carrying
capacity for the environment it lives in. In mathematical terms,
$x_0 = 0$ is an unstable equilibrium, $x_*$ a stable one, and all
nonzero initial data are attracted to $x_*$. Note also that $x(0)
\ge 0$ guarantees $x(t) \ge 0 $ for all $t$, so in the following
we will assume $x \ge 0$. We will always refer to the population
dynamic context for ease of language, but the transposition to
chemical kinetics or other contexts would be immediate.

In this note we want to consider the \emph{stochastic} (Ito)
version of the logistic equation, i.e. \beql{eq:LS} d x \ = \ \(
\a \, x \ - \ \b \, x^2 \) \, d t \ + \ \s (x,t) \, d w \ ; \eeq
here $\a , \b$ are positive real constants, $w = w(t)$ is a Wiener
process, and the diffusion coefficient $\s = \s (x,t)$ takes into
account the magnitude of stochastic effects as well as their
dependence on the level of $x$ and possible explicit dependence on
time $t$.

In particular we want to show that when $\s$ has some specific
form (which is one of the two natural ones in terms of applications, see below)
the equation \eqref{eq:LS} can be explicitly integrated via a
recently proposed technique, i.e. making use of its
\emph{symmetry} properties \cite{\stochsymmref}.

The study of a stochastic version of the logistic equation is of
course not new, and it has been tackled in the literature under
different points of view; see e.g.
\cite{OvaMee,NoiseRev,\appref,\ranref}. As far as we have been
able to determine, however, this has not led to a complete
integration of the model as done here, and of course the analysis has never been on the basis of symmetry considerations and applying the
recently developed theory of \emph{symmetry of stochastic
differential equations} \cite{\stochsymmref} (which generalizes
the classical theory of symmetry of deterministic equations
\cite{\symmref}), which is what we will do here.

\section{The diffusion coefficient}

As mentioned above, in many applications -- in particular in those
of biological or chemical origin -- the variable $x(t)$ represents
the population or the concentration of a given (biological or
chemical) species, and the process modelled by the logistic
equation is a growth/reproduction process in the Biological case,
a reaction issued by the encounter of two types of molecules in
the chemical case.

Thus the fluctuations in the (speed of the) process will arise
from fluctuations in the reproduction process and/or the
environmental conditions, or fluctuations in the concentration of
the chemical species and/or the reaction rate.

In fact, the modeling of these fluctuations will depend on their
nature. Keeping to the Population Dynamic framework, there will be
fluctuations due to the fact that single individuals do not
reproduce at exactly the average rate: we expect their
reproduction rates to have a distribution with given average and
some dispersion, and to be uncorrelated. Thus the diffusion term
in the corresponding stochastic equation \eqref{eq:LS} will be
proportional to $\sqrt{x}$. We also speak of \emph{demographical
noise} \cite{OvaMee}.

But other sources of fluctuations are also present. In particular,
they can depend on environmental conditions, such as availability
of nutrients, temperature, presence of predators, etc. In this
case the fluctuations for different individuals -- or at least for
those living in a given spatial region -- are completely
correlated; in this case biologists speak of \emph{environmental
noise} \cite{OvaMee}. In this case the diffusion term in the
stochastic equation \eqref{eq:LS} is proportional to $x$; in
mathematical terms, this means we have \emph{multiplicative
noise}.

In general both types of noise are present, and we should deal
with a stochastic differential equation of the form
\beql{eq:canmod} d x \ = \ \( \a x \, - \, \b x^2 \) \, d t \ + \
\s_d \, \sqrt{x} \, d w_d \ + \ \s_e \, x \, d w_e \ , \eeq where
$\s_d,\s_e$ are constants and $w_d , w_e$ are Wiener processes
modeling the demographic and the environmental noise
respectively. (Note we are assuming the system does not depend
explicitly on time.) This equation \eqref{eq:canmod} is also
known, in Population Biology, as the \emph{canonical model}
\cite{OvaMee,NoiseRev}.

It should be noted that for large populations the demographical
noise is negligible compared to environmental one\footnote{The
exact balance between the two will of course depend on the
coupling constants $\s_d , \s_e$, so the concept of ``large'' will
be meant in the sense $x \gg (\s_d / \s_e)^2$.}, and in several
cases one is indeed more interested in the effects of
environmental noise. Thus we found investigations of the case
where $\s_d =0$ both in the Theoretical Biology and in the Physics
literature, and this also in more general settings: e.g. for more
general types of noise, for spatially distributed systems, for
interacting populations, etc. \cite{OvaMee,NoiseRev,\appref}.

We want to deal exactly with this case, i.e. a logistic equation
with environmental noise. Thus we will set the diffusion
coefficient $\s$ to be \beq \s (x,t) \ = \ \ga \, x \ , \eeq with
$\ga$ a positive real constant.

With this choice, the equation under study will be \beql{eq:LS0} d
x \ = \ \( \a \, x \ - \ \b \, x^2 \) \, d t \ + \ \ga \, x \, d w
\ . \eeq
Note that the half-line $x \ge 0$ is invariant under this; that is, an initial condition $x(t_0) \ge 0$ will produce a dynamics with $x(t) \ge 0$ for all $t \ge t_0$. In fact, $x=0$ is an impassable barrier not only for the drift, but also for the stochastic term, as the noise coefficient goes to zero for $x=0$; moreover, for the same reason, $x=0$ is a stationary point.

The equation \eqref{eq:LS0} is invariant under the
(two-parameters) scaling group 
\beq x \to \la \, x \ , \ \ \a \to \mu \, \a \ , \ \ \b
\to (\mu/\la) \, \b \ , \ \ \ga \to \sqrt{\mu} \, \ga \ , \ \ t \to (1/\mu) \, t \ . \eeq (Note that the scaling $t \to (\/\mu) t$ will also rescale the Wiener process by a factor $1/\sqrt{\mu}$, hence the need to rescale the $\ga$ parameter.) 

This allows to study a ``universal'' stochastic logistic equation, e.g. with
$\a = \b = 1$, and obtain the description of all special cases
(i.e. given values of $\a$ and $\b$) just by action of this
scaling group. If we do not act on time, $\mu=1$, the scaling reduces to
\beql{eq:scal} x \to \la  x \ , \ \ \b \to \la^{-1} \b \ . \eeq

\section{Symmetry of stochastic equations}
\label{sec:SSE}

Symmetry methods are among the most effective tools in attacking
\emph{deterministic} nonlinear equations \cite{\symmref}. More
recently they have also been applied to the study of
\emph{stochastic} nonlinear differential equations
\cite{\stochsymmref}.

We refer to the literature (see the references listed above, and
in particular the review \cite{GGPR}, for an overview) for the
general results in this context. Here we are interested in the
case of a \emph{scalar} Ito stochastic equation; we will need a
simple classification \cite{GS17} and a theorem originally due to
Kozlov \cite{Koz1,Koz2,Koz3} (see also \cite{GS17,GGPR,GL1,GL2}).
Both of these are briefly recalled in this section for the case of
interest here, i.e. specializing to the case of scalar equations.
If nothing is specified, in the following very sketchy discussion
a stochastic differential equation (SDE) is always meant to
possibly mean a vector one (i.e. a system of coupled scalar
equations). We always consider \emph{ordinary} SDEs
\cite{\sderef}.

\subsection{Admissible maps}

In a stochastic differential equation, we have the time $t$, one
or more Wiener processes $w^k (t)$, and the stochastic processes
described by the SDE itself, $x^i (t)$. So \emph{apriori} we would
consider maps (diffeomorphisms)
$$ \( x,t;w \) \ \to \ \( \wt{x} , \wt{t} ; \wt{w} \) $$ with
$\wt{x} = \wt{x} (x,t;w)$, etc., i.e. general maps in the
$(x,t;w)$ space. When a map leaves the equation under study
invariant, we will say this is a \emph{symmetry} for it.

However, a little thinking shows that such maps are definitely too
general. In particular, $t$ is a smooth variable and it should not
mix with random ones, so we should require $\wt{t} = \wt{t} (t)$;
in other words, we can at most consider reparametrizations of
time. Moreover, we want to map Wiener processes into Wiener
processes. Albeit a scalar factor could be absorbed into the
diffusion matrix, we need to preserve independence of the
different Wiener processes. This leads to consider maps of the
form $\wt{w}^i = R^i_{\ j} w^j$, with $R$ a constant conformal
matrix (that is, belonging to the conformal group). We have thus
identified the \emph{admissible maps} in $(x,t;w)$ space.

Finally, albeit we could always consider general maps (within the
class identified above), actually we know that considering
\emph{infinitesimal transformations} will be specially productive.
Thus we consider generators for such transformations, in the form
\beql{eq:X0} X \ = \ \vphi^i \, \frac{\pa}{\pa x^i} \ + \ \tau \,
\frac{\pa}{\pa t} \ + \ h^k \, \frac{\pa}{\pa w^k} \ . \eeq If a
SDE $\mathcal{E}$ is invariant under the action of such a vector
field (acting in the $(x,t;w)$-space), we will say $X$ is a
(Lie-point) \emph{symmetry generator} for $\mathcal{E}$. By a
standard abuse of terminology, we will also say, for short, that
$X$ is a \emph{symmetry} for $\mathcal{E}$.

The discussion above identified admissible maps; when we translate
this into the case of infinitesimal maps, i.e. vector fields in
the form \eqref{eq:X0}, this means we should require $\tau = \tau
(t)$ and $h^k = R^k_{\ m} w^m$. Thus finally we will consider
vector fields of the form \beql{eq:XG} X \ = \ \vphi^i (x,t;w) \,
\frac{\pa}{\pa x^i} \ + \ \tau (t) \, \frac{\pa}{\pa t} \ + \ \(
R^k_{\ m} w^m \) \, \frac{\pa}{\pa w^k} \ . \eeq These will be
dubbed \emph{admissible} vector fields, and these will be the only
class of vector fields to be considered as candidates to be
symmetries of the SDE under study.

The reader can consult \cite{GS17,GGPR} for a more detailed
discussion of admissible maps and vector fields.

\subsection{Classification of symmetries}

The admissible vector fields \eqref{eq:XG} will induce an action
on the space of Ito SDEs \beql{eq:ItoG} d x^i \ = \ f^i (x,t) \, d
t \ + \ \s^i_{\ k} (x,t) \, d w^k \eeq (i.e. of their coefficients
$f$, $\s$) and map an equation $\mathcal{E}$ into a, generally
different, equation $\mathcal{F} = \mathcal{E} + \eps \delta
\mathcal{E}$. As mentioned above, when $\mathcal{F} =
\mathcal{E}$, i.e. when $\delta \mathcal{E} = 0$, we say that $X$
is a symmetry for $\mathcal{E}$.

Depending on the special features of the vector field $X$, we can
have different types of symmetries. In particular:

\begin{itemize}

\item If $\tau = 0$, we have a \emph{simple} symmetry;

\item If $R=0$, we have a \emph{standard} symmetry; standard
symmetries can be \emph{deterministic} if $\vphi^i$ do not depend
on $w$, or \emph{random} if (at least one of) the $\vphi^i$ does
depend on (at least one of) the $w^k$;

\item If $R\not=0$, hence $X$ acts effectively on the $w^k$ variables, then we have a \emph{W-symmetry}. \footnote{Any standard symmetry can also be considered, for ease of discussion, a \emph{trivial} W-symmetry, with $R=0$.}

\end{itemize}

The Kozlov theory of symmetry integration of SDEs (and
generalizations) makes use only of \emph{simple} symmetries
\cite{Koz1,Koz2,Koz3}, and hence our attention will be focused on
these.

Random symmetries were introduced in \cite{GS17} and studied in a
number of other papers \cite{GL1,GL2,Koz18a,Koz18b}; W-symmetries
were introduced in \cite{GS17} and further studied in \cite{GW18}.

Finally, a simple but most relevant remark: in changing variables,
vector fields transform under the familiar chain rule, but
stochastic processes and Ito equations transform under the Ito
rule. Thus it is not at all obvious, apriori, that symmetries will
survive a change of variables. It turns out that -- \emph{in the
case of admissible vector fields} -- this is precisely what
happens, namely symmetries are preserved under change of variables
and are thus an intrinsic feature of the stochastic process, not
conditional upon its coordinate description. See \cite{GL1} for a
detailed discussion.

\subsection{Determination of symmetries}

By studying how the $X$ action modifies the coefficients $f^i$,
$\s^i_{\ k}$ of the general Ito equation \eqref{eq:ItoG} one finds
the relation which must exist between these and the coefficients
of the vector field $X$ for the equation to be invariant. These
relations go under the name of \emph{determining equations} for
the symmetries of Ito equations. These are discussed in the
general case in the literature, see e.g. \cite{GS17,GGPR,GW18};
here we will just consider the case of interest here, i.e. that of
\emph{scalar} equations. (This allows for a simpler notation
compared with the general one.)

Simple standard symmetries of a given (scalar) Ito equation
\beql{eq:Ito} d x \ = \ f(x,t) \, d t \ + \ \s (x,t) \, d w \ ,
\eeq i.e. symmetry vector fields of the form \beql{eq:Xstand} X \
= \ \vphi^i (x,t;w) \, \pa_i \ , \eeq are determined as solution
to the determining equations (see e.g. \cite{GS17,GGPR} for their
derivation)
\begin{eqnarray}
\vphi_t \ + \ f \, \vphi_x \ - \ \vphi \, f_x \ = \ - \, \frac12
\Delta ( \vphi) \ , \label{eq:IDE1} \\
\vphi_w \ + \ \s \, \vphi_x \ - \ \vphi \, \s_x \ = \ 0 .
\label{eq:IDE2}
\end{eqnarray}

In this formula we have introduced the \emph{Ito Laplacian}
$\Delta$; when there are only a single $x$ and a single $w$, as in
our case, this is defined as \beql{eq:Delta} \Delta \ = \
\frac{\pa^2}{\pa w^2} \ + \ 2 \, \s \, \frac{\pa^2}{\pa x \pa w} \
+ \ \s^2 \, \frac{\pa^2}{\pa x^2} \ . \eeq Here $\s = dx/dw$ is
the same diffusion coefficient appearing in \eqref{eq:Ito}.

As for W-symmetries, i.e. in the  scalar case symmetry vector
fields of the form \beql{eq:XW} X \ = \ \vphi (x,t;w) \, \pa_x \ + \ R \,
w \, \pa_w \ , \eeq the determining equations for these (derived
in \cite{GW18}) are
\begin{eqnarray}
\vphi_t \ + \ f \, \vphi_x \ - \ \vphi \, f_x  &=& - \frac12
\Delta (\vphi) \ ; \label{eq:deteqw1} \\
\vphi_w \ + \ \s \, \vphi_x \ - \ \vphi \, \s_x &=& R \, \s \ .
\label{eq:deteqw2}
\end{eqnarray}
Note the first equation is the same as for standard symmetries,
while the second one is different from that case.

Note also that the equations are \emph{linear} in $\vphi$; thus --
among other consequences -- we will always have an arbitrary
multiplicative constant $c$ in any solution (hence in any
symmetry); this is unessential and will be set to the most
convenient value, typically $c=1$.

As already mentioned, we refer e.g. to \cite{GS17,GGPR,GW18} for
details on the derivation of these determining equations and for
their version in arbitrary dimension.

\subsection{Symmetry and symmetry adapted variables}
\label{sec:SAV}

The Kozlov theory shows that if a scalar SDE admits a simple
symmetry $X$, then it can be explicitly integrated by passing to
new, symmetry-adapted, variables \cite{Koz1,Koz2,Koz3}. The
converse is also true, i.e. if a scalar SDE can be integrated in
this way, then it necessarily admits a simple symmetry \cite{GL2}.
In the case of general (multi-dimensional) equations, a symmetry
corresponds to reduction of the dimensionality of the equation.

What is more relevant, is that the theory is constructive. In
other words, if we identify a symmetry (by solving the determining
equations), then the needed change of variables can be explicitly
built by a simple general formula.

It is convenient to discuss separately the cases of standard
symmetries and of W-symmetries.

\subsubsection{Standard symmetries}
\label{sec:SAVsta}

The basic result for the use of standard symmetries was provided
by Kozlov \cite{Koz1,Koz2,Koz3}. Here we quote it from
\cite{GW18}, see Proposition 3 in there.

\medskip\noindent
{\bf Proposition 1.} {\it Let the scalar Ito equation
\eqref{eq:Ito} admit the simple standard vector field $ X  =
\vphi (x,t;w) \pa_x$ as a Lie-point symmetry; then by passing to the
new variable \beql{eq:Phi} y \ = \ \Phi (x,t;w) \ = \ \int
\frac{1}{\vphi(x,t;w)} \ d x \eeq the equation is in general
mapped into \beql{eq:newgen} d y \ = \ F (t;w) \, d t \ + \ S(t;w)
\, d w \eeq and hence is readily integrated as \beql{eq:newgensol}
y(t) \ = \ y(t_0) \ + \ \int_{t_0}^t F [ t , w(t)] \, d t \ + \
\int_{t_0}^t S [ t , w(t)] \, d w(t) \ . \eeq}
\bigskip

In order to provide complete information, we also quote the
following result, which establishes when the transformed (and
integrable) equation \eqref{eq:newgen} is actually of Ito type
\cite{GS17,GL2,GW18}.

\medskip\noindent
{\bf Lemma 1.} {\it In the setting of Proposition 1, the equation
\eqref{eq:newgen} is in Ito form,
$d y  =  F (t)  d t  +  S(t)  d w$,
%\beql{eq:newito} d y \ = \ F (t) \, d t \ + \ S(t) \, d w \ , \eeq
%and hence its solution is \beql{eq:newitosol} y(t) \ = \ y(t_0) \ + \
%\int_{t_0}^t F ( t ) \, d t \ + \ \int_{t_0}^t S ( t ) \, d w(t) \ , \eeq
if and only if the functions
$f(x,t)$, $\s(x,t)$ and $\psi(x,t;w) := \pa_w [\vphi(x,t;w)]^{-1}$
%\beql{eq:defpsi} \psi(x,t;w) \ := \ \frac{\pa}{\pa w} \, \( \frac{1}{\vphi(x,t;w)} \) \eeq
satisfy the relation \beql{eq:psirel} \s \, \psi_t \ + \ \s_t \, \psi \ = \ f
\, \psi_w \ + \ \frac12 \, \( \s \, \psi_{ww} \ + \ \s^2 \,
\psi_{xw} \) \ . \eeq }
\bigskip

It should be stressed that \eqref{eq:newgensol}
%and \eqref{eq:newitosol} make
makes use of Ito integrals (alongside standard ones), and that
even when \eqref{eq:psirel} is not satisfied, the resulting
equation for the new variable $y$ is readily integrated.\footnote{We also stress that,
conversely, if the Ito equation \eqref{eq:Ito} is reducible to the
integrable form \eqref{eq:newgen} by a simple random change of
variables $y = \Phi (x,t;w)$ then necessarily \eqref{eq:Ito}
admits $X = \left[ \Phi_x (x,t,w) \right]^{-1} \pa_x := \varphi (x,t,w)
\pa_x$ as a symmetry vector field, and when \eqref{eq:newgen} is
actually of Ito form then \eqref{eq:psirel} is satisfied with
$\psi = \pa_w (1/\vphi)$ \cite{GL2}.}

We also note, in connection with Lemma 1, that if the map does not
mix the $(x,t)$ and the $w$ variables -- in which case we speak of
a \emph{split} W-map -- then we are guaranteed the transformed
equation is again of Ito type \cite{GW18}. In terms of the vector
field \eqref{eq:Xstand}, this means that $\vphi_w = 0$.

\subsubsection{W symmetries}
\label{sec:SAVW}

In the case of W-symmetries, as discussed in detail in
\cite{GW18}, once a W-symmetry, call it $X$, has been determined
our integration strategy is in principles the same as for standard
symmetries, i.e. passing to \emph{symmetry adapted coordinates};
but it is implemented in a slightly different way.

It turns out that in the present case, i.e. for the stochastic
logistic equation \eqref{eq:LS0}, no nontrivial W-symmetries are
present. Thus we will not discuss integration under W-symmetries
(for this, see \cite{GW18}).

\section{Symmetry analysis of the stochastic logistic equation}
\label{sec:SASLE}

We will now determine the symmetries of the stochastic logistic
equation. We will first study the simplest case of standard
symmetries, and then the case of W-symmetries.

\subsection{Standard symmetries}

We start by considering \emph{standard} symmetries of the
stochastic logistic equation \eqref{eq:LS0}. Now the determining
equations \eqref{eq:IDE1} and \eqref{eq:IDE2} have a more specific
form; in particular the second one, \eqref{eq:IDE2}, reads
$$ \vphi_w \ + \ \ga \, \( x \, \vphi_x \ - \ \vphi \) \ = \ 0 \ .
$$ This yields
$$ \vphi (x,t;w) \ = \ x \ q (t,z) \ , $$
having defined
$$ z \ := \ w \ - \ \frac{\log (x)}{\ga} \ . $$
With this expression for $\vphi$ (and the given expressions for
$f$ and $\s$), the first determining equation \eqref{eq:IDE1}
reads
$$ \left[ q_t \, + \, \( \frac{\ga}{2} - \frac{\a}{\ga} \) \, q_z \right] \ x \ + \
\b \, \( q \, + \, \frac{1}{\ga} \, q_z \) \ x^2 \ = \ 0 \ . $$
The coefficients of $x$ and of $x^2$ must vanish separately, and
thus we have two equations for $q$. The one stemming from the
coefficient of $x^2$ yields
$$ q(t,z) \ = \ \exp [- \ga z] \ r(t) \ ; $$
plugging this into the $x$ coefficient we get
$$ e^{- \ga z} \ \left[ r' \ + \ (\a - \ga^2/2 ) \, r \right] \ = \ 0 $$
and therefore (here $c_1$ is an arbitrary constant)
$$ r(t) \ = \ c_1 \ \exp \left[ - \( \a - \ga^2/2 \) t \right] \ . $$

Thus in the end we have one simple standard symmetry; setting the
unessential constant $c_1$ to unity and introducing (for ease of
notation here and in the following) the new constant \beql{eq:Apm}
A \ := \ \a \ - \ \ga^2/2 \ , \eeq the symmetry is identified by
\beql{eq:stasym} \vphi(x,t;w) \ = \ x^2 \ \exp
\left[ - A \, t \ - \ \ga \, w \right] \ . \eeq

This is a genuinely random standard symmetry (our computation also
shows that there are no simple \emph{deterministic} standard
symmetries); one can easily check that \eqref{eq:psirel} is
\emph{not} satisfied in this case, so on the basis of Lemma 1 one
knows that the transformed equation will not be of Ito type (this
will be confirmed by our explicit computations below).

\subsection{W-symmetries}

Computations are performed along the same scheme when we search for
W-symmetries. Now the determining equations are \eqref{eq:deteqw1}
and \eqref{eq:deteqw2}; the second of these reads
$$ \vphi_w \ - \ \ga \, \vphi \ + \ \ga \, x \, \( \vphi_x \, - \, R
  \) \ = \ 0 \ . $$
This yields, writing again $z = w - \ga^{-1} \log (x)$,
$$ \vphi (x,t;w) \ = \ x \ \left[ R \, \log (x) \ + \ q (t ,z) \right] \ . $$
Plugging this into \eqref{eq:deteqw1} we get
$$ \mathcal{E}_1 \, x \ + \ \mathcal{E}_2 \, x^2 \ + \ \mathcal{E}_3 \,
x^2 \, \log (x)
\ = \ 0 \ ; $$
here we have written
\begin{eqnarray*}
\mathcal{E}_1 &=& q_t \ + \ \( \frac{\ga}{2} - \frac{\a}{\ga} \) \, q_z \ + \
\( \a + \frac{\ga^2}{2} \) \, R \ ; \\
\mathcal{E}_2 &=& \frac{\b}{\ga} \, q_z \ + \ \b \, q \ - \ \b \,
R \ ; \\
\mathcal{E}_3 &=& \b \, R \ . \end{eqnarray*}

Again the coefficients of different powers of $x$ must vanish
separately and we get three equations $\mathcal{E}_j = 0$ for
$q(t,z)$. But, the equation $\mathcal{E}_3=0$ yields that either
$\b = 0$ or $R =0$. The case $\b=0$ is excluded, as we assumed
both $\a$ and $\b$ are positive real constants (and for $\b = 0$
we would indeed just have a linear SDE), so it must be $R=0$. This
means we only have \emph{trivial} W-symmetries, i.e. the standard
symmetries considered above.

\section{Integration of the stochastic logistic equation}
\label{sec:ISLE}

In this Section we will use the tools described in Section
\ref{sec:SSE}, and in particular in Section \ref{sec:SAV},
together with the results of the symmetry analysis conducted in
Section \ref{sec:SASLE}, in order to integrate the stochastic
logistic equation \eqref{eq:LS0}.

The result of the previous Section \ref{sec:SASLE} also shows we
only deal with standard symmetries.

According to the Kozlov prescription, see Proposition 1, once we
have determined a simple standard symmetry $X = \vphi \pa_x$ of
the Ito equation we should pass to the new variable
\beql{eq:xtoy0} y \ = \ \Phi (x,t;w) \ = \ \int \frac{1}{\vphi
(x,t;w)} \ d x \ ; \eeq with the symmetry determined above, see
\eqref{eq:stasym}, this means we should consider \beql{eq:xtoy} y
\ = \ - \ \frac{\exp \left[  A \, t \ + \ \ga \, w  \right]}{x} \ = \
\Phi (x,t;w) \ . \eeq The inverse change of variables is of course
\beql{eq:ytox} x \ = \ - \ \frac{\exp \left[  A \, t \ + \ \ga \,
w  \right]}{y} \ . \eeq

Note that in \eqref{eq:xtoy} we have implemented \eqref{eq:xtoy0}
and set the integration constant, which is actually an arbitrary
function of $t$ and $w$, to zero.
Note also that the dynamically invariant half-line $x \ge 0$ is
mapped into $y \le 0$, and {\it vice-versa}.

The evolution of the $y$ variable is described by
\beq 
d y \ = \ \frac{\pa \Phi}{\pa t} \, d t \ + \ \frac{\pa \Phi}{\pa x} \, d x \ + \
\frac{\pa \Phi}{\pa w} \, d w \ + \ \frac12 \ \Delta (\Phi) \, d t  \ = \ 
\frac{\Phi}{x^2} \ \left[ - x \, dx \ + \ \a
\, x^2 \, d t \ + \ \ga \, x^2 \, d w  \right] \ ;
\eeq in conclusion we get, using \eqref{eq:LS0} to
express $dx$, regrouping terms, and expressing then $x$ as in
\eqref{eq:ytox}, \beql{eq:dy} d y \ = \  - \, \b \ \exp \left[  A
\, t \ + \ \ga \, w  \right] \ d t \ . \eeq Note that the half-line
$y \le 0$ is invariant under this. This equation also inherits the
scaling \eqref{eq:scal}: more precisely, it is invariant under
\beql{eq:yscal} y \to \la \, y \ , \ \ \b \to \la \, \b \ . \eeq

We are thus reduced to an equation (not in Ito form, as expected)
which obviously admits $\pa_y$ as a symmetry, and is readily
integrated to yield \beql{eq:dysol} y (t) \ = \ y(t_0) \ - \ \b \
\int_{t_0}^t \exp \left[ A \, \tau \ + \ \ga \, w (\tau)  \right] \
d \tau \ . \eeq

\medskip\noindent
{\bf Remark.} As mentioned above we would have in full generality
$$ y \ = \ - \ \frac{\exp \left[  A \, t \ + \ \ga \, w \right]}{x} \ + \
\rho (t,w) \ . $$ In this case we would get, with the same
computation,
\beql{eq:dygen} d y \ = \ \left[ - \b \, \exp [ A t + \ga w] \ + \
\rho_t \ + \ \frac12 \, \rho_{ww} \right] \, d t \ + \
\rho_w \, d w \ . \eeq This is again
immediately integrable, with a slightly more complex explicit
formula:
\begin{eqnarray*}
y (t) &=& y(t_0) \ - \ \b \ \int_{t_0}^t \exp \left[  A \, \tau \
+ \ \ga \, w (\tau)  \right] \ d \tau \\ & & \ + \ \int_{t_0}^t \(
\rho_\tau [\tau ,w(\tau )] \ + \
\frac12 \, \rho_{ww} [\tau ,w(\tau )] \) \, d \tau  
\ + \ \int_{t_0}^t \rho_w [\tau , w(\tau )] \ d w (\tau) \ .
\end{eqnarray*} It is natural to wonder if a suitable choice of
the function $\rho (t,w)$ could make the coefficients of $dt$ and
of $dw$ in the SDE \eqref{eq:dygen} describing the evolution of
$y$ independent of $w$, i.e. if we could obtain in this way an Ito
equation. This is not the case, basically due to the fact $w$
appears in an exponential term \footnote{More precisely, if we
require that the coefficient $F (t,w)$ of $dt$ in \eqref{eq:dygen}
is independent of $w$, i.e. study the equation $\pa_w F = 0$, we
obtain the unique solution $\rho (t,w) = (\b / \a) \exp [(\a -
\ga^2/2) t + \ga w] + c_1 \exp [\ga w - (\ga^2/2) t]  $, which
yields $F=0$.}. Hence there is no advantage in considering a
nonzero $\rho$ and a more complex expression for $\Phi$ than the
one dealt with above.

\section{Numerical experiments}

Our previous discussion provides a mathematically rigorous
construction of the general solution for the stochastic logistic
equation. %\footnote{Sometimes in the mathematical literature such
%a solution, providing explicitly the process $x(t)$ for each
%realization of the driving process $w(t)$ is called a
%\emph{strong} solution of the SDE.}

The reader less inclined to mathematically abstract discussions
may be willing to have a check that our constructions is indeed
reaching its task, e.g. by numerically comparing the solution
obtained in this way with a direct numerical solution of the
stochastic logistic equation \eqref{eq:LS0} for the same
realization of the driving Wiener process $w(t)$.

Such a request would be even more justified considering that --
quite surprisingly -- in the recent literature devoted to symmetry
of SDEs and integration of the latter by symmetry method such
``numerical check'' appear to be completely absent (also in
publications by the present author).

We have thus ran a ``numerical experiment'' consisting of the
following procedure (repeated over a number of realizations of the
driving process).

\begin{enumerate}

\item Generate, by means of a random number generator, a sequence
of normally distributed $(\de w) (k)$ (for $k = 0, ... ,
k_{\mathrm{max}} = k_M $); store these.

\item Using the stored values of $(\de w)(k)$, build a
discrete-time Wiener process (time step $\de t$) setting $w[0]=0$,
$w[(k+1)] = w [ k ]  +  (\de w) (k) \de t$; here $w[k]$ represents
the value taken by $w(t)$ at time $t_k = k (\de t)$. Store this.

\item Numerically integrate \eqref{eq:LS0}, again using the stored
$(\de w)(k)$, by setting $x(0) = x_0$, $ x[ (k+1) ] =  x [ k ] + (
\a x[k] - \b x[k]^2 )  dt  +  \ga  x[k] (\de w)[k]$, and store
these. Here $x[k]$ represents the value taken by $x(t) $ at time
$t_k = k (\de t)$, for the given realization of $w(t)$. The values
$x[k]$ represent a \emph{bona fide} direct numerical solution of
our stochastic equation, with the approximation resulting from the
finite size of the time step $(\de t)$. \footnote{We stress we are
here using a very basic Euler first order integration scheme, thus
have to expect rather poor precision in our numerical results. One
could use more refined integration schemes -- see e.g. \cite{Doe}
-- but the point here is just to have a reference numerical
solution to compare our exact solution with.}

\item Use the map \eqref{eq:xtoy} to determine $y(0) = y_0$
corresponding to the assigned initial value $x_0$. Using the
stored values of $w[k]$ (i.e. of $w(t)$ for the given
realization), build $y(t)$ by means of \eqref{eq:dy}, i.e. setting
$y[0] = y_0$ and $ y[k+1] = y[0] - \b  \exp ( A \, t_k \ + \ \ga
\, w[k] )  \de t$; here again $y[k]$ represents the value taken by
$y(t)$ at time $t_k$. The values $y[k]$ are stored and represent a
\emph{bona fide} direct numerical solution of our equivalent
stochastic equation \eqref{eq:dy}, i.e. of \eqref{eq:dysol}; again
with the to approximation due to finite size of $\de t$.

\item Use now the inverse map \eqref{eq:ytox} to generate from the
values stored in $Y$ -- i.e. from $y(t)$ -- values $\^x[k]$ which
represent the values taken by a stochastic process $\^x (t)$ at
time $t_k$.

\item If our procedure is correct, the stochastic process $\^x
(t)$ is the solution to the equation \eqref{eq:LS0} for the given
realization of $w(t)$, i.e. should correspond to $x(t)$ computed
directly before. Thus we compare the strings $x[k]$ and $\^x[k]$.

\end{enumerate}

We stress that one expects some disagreement to be present due to
the numerical errors and above all to the finite size of the
considered time step of the numerical integration. It would be
possible to reduce these by considering a more refined numerical
integration scheme, but this is not of interest here: we only want
to check that our procedure is correct in that it does indeed
produce a solution to the equation under study, and this is
clearly shown by our rough numerical computations.

\begin{figure}
\begin{tabular}{cc}
  \includegraphics[width=180pt]{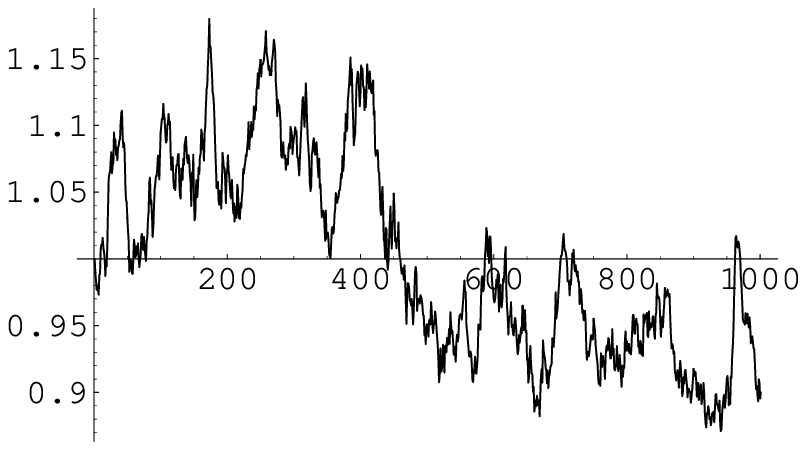} &
  \includegraphics[width=180pt]{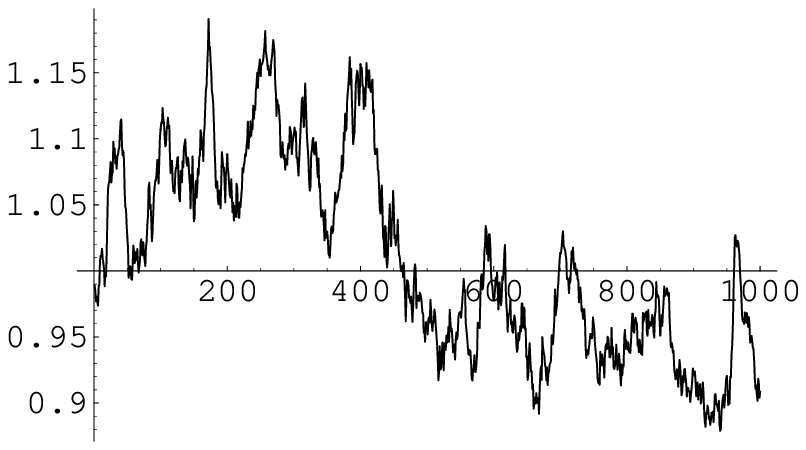}\\
 (a) & (b) \\
  & \\
  \includegraphics[width=180pt]{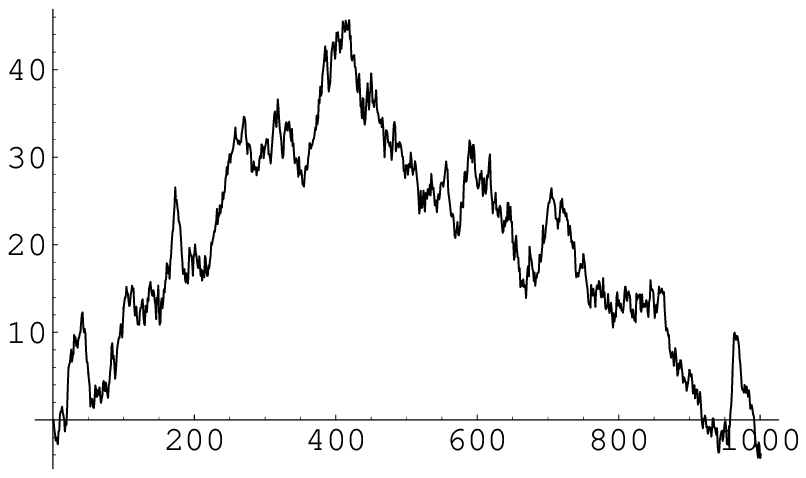} &
  \includegraphics[width=180pt]{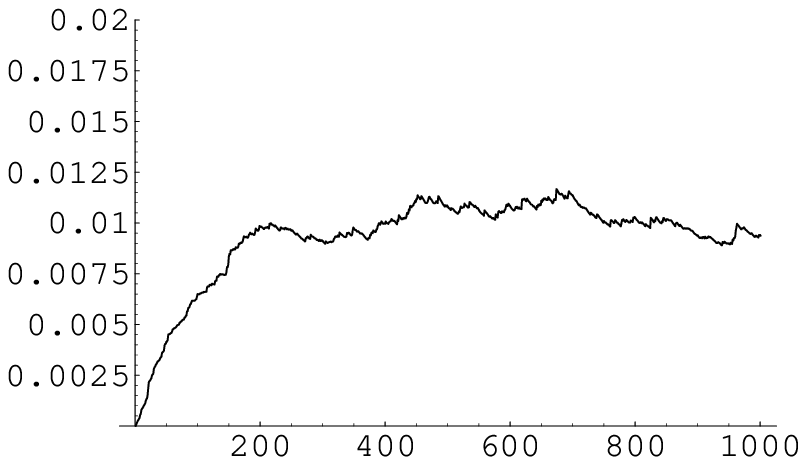}\\
 (c) & (d) \end{tabular} \\
 \caption{For $\a=\b=1$, $\ga = 0.01$, $x_0 = \a/\b$ (the equilibrium solution
 for zero noise) and a specific realization of $w(t)$ we show:
 (a) the direct numerical solution $x(t)$;
 (b) the solution $\^x(t)$ obtained by our procedure;
 (c) the realization of the driving stochastic process $w(t)$;
 (d) the relative error $|\^x (t) - x(t)|/x(t)$. The numerical integrations are performed over 1.000 steps.} \label{fig1}
\end{figure}

\begin{figure}
\begin{tabular}{cc}
  \includegraphics[width=180pt]{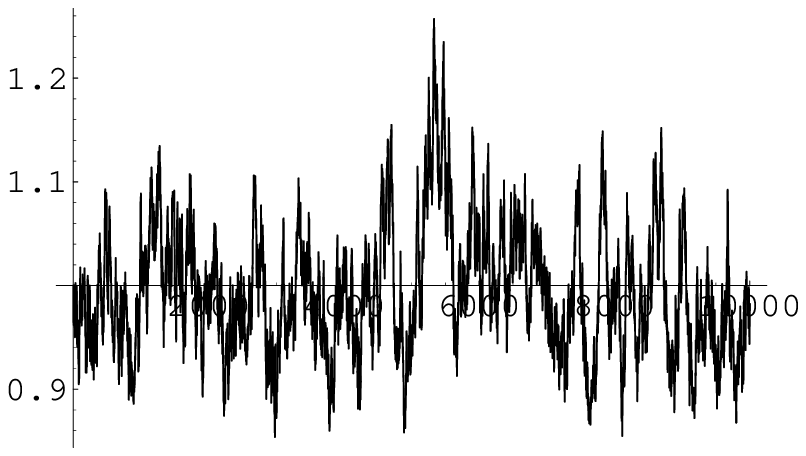} &
  \includegraphics[width=180pt]{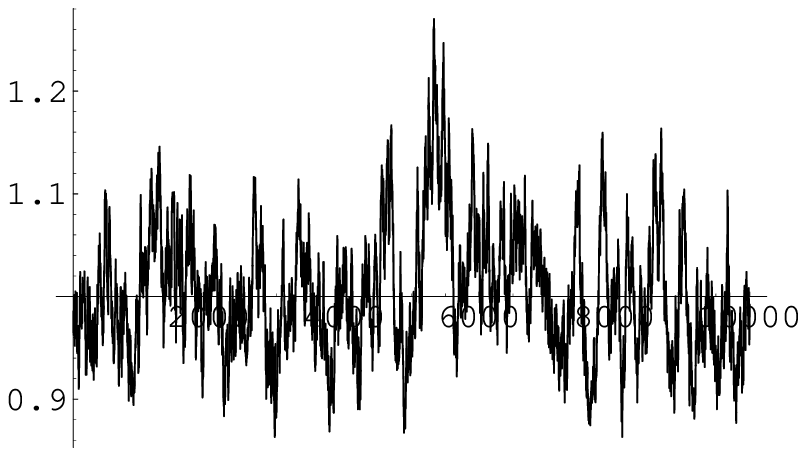}\\
 (a) & (b) \\
  & \\
  \includegraphics[width=180pt]{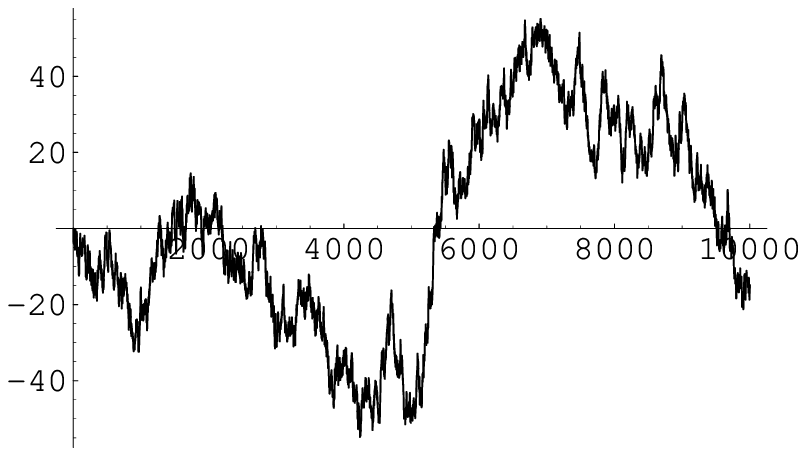} &
  \includegraphics[width=180pt]{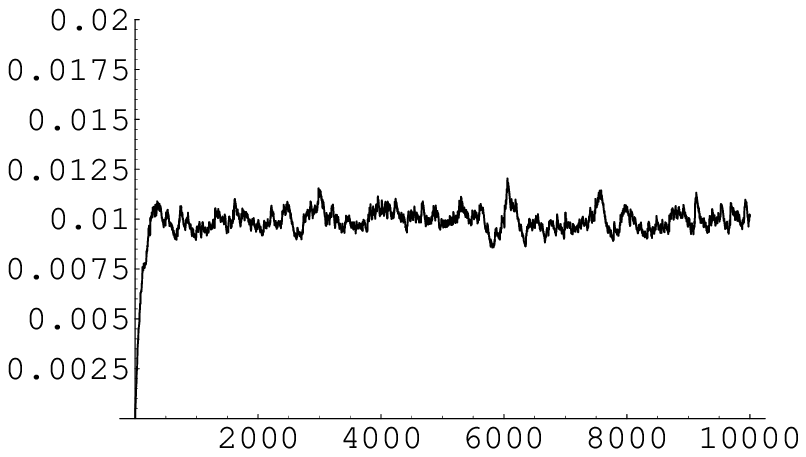}\\
 (c) & (d) \end{tabular} \\
  \caption{For the same parameters values as in Fig.\ref{fig1}
  we show  again: (a) $x(t)$; (b) $\^x(t)$; (c) $w(t)$;
  (d) $|\^x (t) - x(t)|/x(t)$. Here the numerical integrations are performed over 10.000 steps.} \label{fig2}
\end{figure}

The results of these numerical experiments are shown (for two
given realizations of the driving process; we have of course
conducted many runs) in Figures \ref{fig1} and \ref{fig2} (see
captions there for details), and confirm that indeed our procedure
provides a correct solution to the stochastic logistic equation
for each realization of the driving Wiener process (the
disagreement between $x$ and $\^x$ remains around one percent
over 10.000 time steps).

\section{Conclusions}

We have considered the stochastic logistic equation
\eqref{eq:LS0}. We have shown that in order to integrate it, i.e.
to provide $x(t)$ for given $x(0)$ and \emph{for each realization
of the Wiener process $w(t)$} one can pass to consider the random
variable $y(t)$ defined by \eqref{eq:xtoy}. This evolves according
to \eqref{eq:dy} and is therefore given by \eqref{eq:dysol}. Going
back to the original variable amounts to applying \eqref{eq:ytox}.

This allows to obtain much more precise information about
solutions to \eqref{eq:LS0}; in particular one can obtain
information \emph{for each realization} of the driving process
$w(t)$.

We have also checked by numerical computations for different given
realization of the driving process that our procedure does indeed
provide a solution to the stochastic logistic equation.

\section*{Acknowledgements}

\noindent Most of this work was performed over a stay at SMRI,
providing as usual a relaxed atmosphere and convenient working
conditions. %Useful discussion with L. Peliti (the deputy Director of SMRI) are gratefully acknowledged.
My work is also supported by GNFM-INdAM.

\end{document}